# Interfacial Control of Dzyaloshinskii–Moriya Interaction in Heavy Metal/Ferromagnetic Metal Thin Film Heterostructures

Xin Ma[1], Guoqiang Yu[2], Xiang Li[2], Tao Wang[3], Di Wu[2], Kevin S. Olsson[1], Zhaodong Chu[1], Kyongmo An[1], John Q. Xiao[3], Kang L. Wang[2] and Xiaoqin Li[1],*

[1]Department of Physics, The University of Texas at Austin, Austin, Texas 78712, USA
[2]Department of Electrical Engineering, University of California, Los Angeles, California 90095, USA
[3]Department of Physics and Astronomy, University of Delaware, Newark, Delaware 19716, USA

*Email address: elaineli@physics.utexas.edu

The interfacial Dzyaloshinskii–Moriya Interaction (DMI) in ultrathin magnetic thin film heterostructures provides a new approach for controlling spin textures on mesoscopic length scales. Here we investigate the dependence of the interfacial DMI constant $D$ on a Pt wedge insertion layer in Ta/CoFeB/Pt(wedge)/MgO thin films by observing the asymmetric spin wave dispersion using Brillouin light scattering. Continuous tuning of $D$ by more than a factor of three is realized by inserting less than one monolayer of Pt. The observations provide new insights for designing magnetic thin film heterostructures with tailored $D$ for controlling skyrmions and magnetic domain wall chirality and dynamics.

## I. INTRODUCTION

The Dzyaloshinskii–Moriya interaction (DMI) refers to anti-symmetric exchange interaction that promotes canted instead of the parallel or anti-parallel spin alignments. Understanding and controlling the DMI may facilitate the design of the next-generation magnetic memory and logic devices based on chiral magnetic domain walls [1-5] and skyrmions[6-11]. Fast current-induced magnetic domain wall motion has been recently demonstrated via the combination of a chiral domain wall structure and spin-orbit torque, where the direction and the speed of domain wall motion depend on both the sign and the strength of the DMI and the spin-orbit torque[1, 2]. Moreover, the DMI is responsible for establishing and controlling the sizes of magnetic skyrmions, which are topologically protected vortex- or hedgehog-like spin structures[7]. These small chiral spin textures show promises in future spintronic applications due to their unique properties including driven propagation by ultralow current densities[8, 10, 11] and rewritability by spin-polarized currents[6].

The magnitude of the DMI can be significant at the interface between a ferromagnetic metal (FM) and a nonmagnetic heavy metal (HM) possessing strong spin-orbit coupling. The enhanced DMI originates from the three-site indirect exchange interaction[12] and helps to stabilize the skyrmion bubble phase at room temperature[9, 13-16]. Moreover, such an interfacial DMI in multilayer heterostructures strongly depends on material composition, layer sequence, and interface quality among other factors[1, 4, 17-22]. This broad parameter space offers unique opportunities to elucidate the underlying physical origin of the interfacial DMI.

In this Rapid communication, we demonstrate effective interfacial control of the DMI in annealed Ta/CoFeB/(Pt)/MgO multilayers. The sign and magnitude of the DMI constant $D$ are obtained from the asymmetric spin wave dispersion probed with Brillouin light scattering (BLS). Continuous tuning of $D$ by more than a factor of three is realized by inserting a Pt-wedge with nominal thickness of less than one monolayer between the CoFeB and MgO layers. Both Ta/CoFeB and CoFeB/Pt interfaces contribute to the overall DMI but with different signs of $D$. The thicker region of the Pt-wedge shows a larger magnitude of $D$ at the CoFeB/Pt interface, leading a smaller net $D$ value in the multilayer structure. This somewhat surprising finding of strong DMI strength modulation via an ultrathin Pt insertion layer together with magnetic isotropy change in this multilayer heterostructure provides new approaches for controlling skyrmions and magnetic domain walls for new and emerging spintronic applications.

## II. Experiments

A series of Ta(5)/Co$_{20}$Fe$_{60}$B$_{20}$(1)/Pt($t_{Pt}$)/MgO(2)/Ta(2) thin films were deposited by magnetron sputtering at room temperature on thermally oxidized silicon substrates as shown in Fig. 1a, where the numbers in parentheses represent the nominal layer thicknesses in nanometers. For one of the multilayer structures, the wedge-shape Pt-insert layer between the CoFeB and MgO layers has the thickness variation from 0.11 to 0.29 nm across the 5-cm-long sample [13]. Following the deposition, all multilayer structures were annealed at 250 ºC for 30 minutes to enhance the perpendicular magnetic anisotropy (PMA).

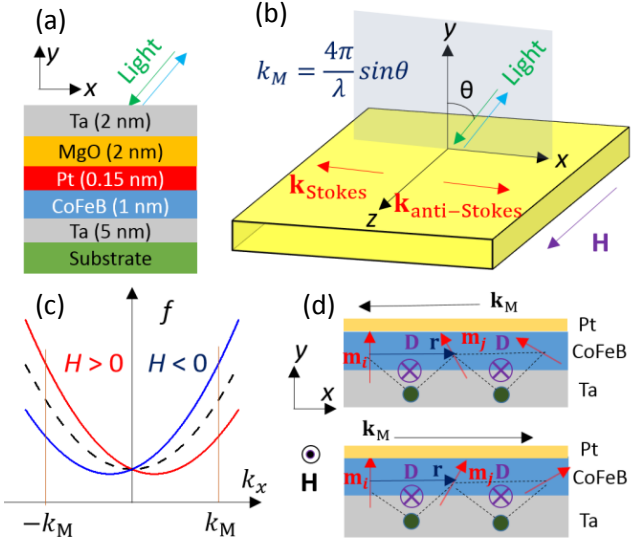

Fig. 1. Schematics of (a) multilayer thin film and (b) BLS measurement geometry. (c) Spin wave dispersions without (dashed) and with (solid) the interfacial DMI $D > 0$. (d) Schematics of spatial chiralities of DE spin waves propagating in opposite directions. Canted red arrows $\mathbf{m}_i$ and $\mathbf{m}_j$ refer to the spin rotation in the $x$-$y$ plane away from the external field direction $+z$ at different atomic sites $i$ and $j$, respectively. The DMI vector $\mathbf{D}$ points into the paper plane ($D > 0$) for the two spins with spatial displacement $\mathbf{r}$. The green dots represent the Ta atoms, and the dashed lines indicate the three-site indirect exchange.

BLS measurements were performed to investigate the asymmetric spin wave dispersion caused by the interfacial DMI. Figure 1b shows the geometry of BLS experiment, where a magnetic field $\mathbf{H}$ was applied along the $z$ axis in all measurements. An s-polarized laser beam was incident on the sample, and the p-polarized component of the backscattered light was collected and sent to a Sandercock-type multipass tandem Fabry-Perot interferometer. In order to reduce the uncertainty in $k_M$, a lens with a focal length of 10 cm was used to focus the incidence light and to collect the scattering signal. In addition, an aperture with 5 mm diameter was placed near the lens as a spatial filter. In the light scattering process, the total momentum is conserved in the plane of the thin film. As a result, the Stokes (anti-Stokes) peaks in BLS spectra correspond to the creation (annihilation) of magnons with momentum $k_M = \frac{4\pi}{\lambda}\sin\theta$ along $-x$ ($+x$) direction as illustrated in Fig. 1b, where $\lambda = 532\ nm$ is the laser wavelength, and $\theta$ refers to the light incident angle. Owing to the presence of the DMI, shifts of spin wave dispersions are introduced as displayed in Fig. 1c. Such shifts are directly reflected in the frequency difference between the Stokes ($-k_M$) and anti-Stokes ($k_M$) peaks in BLS spectrum[17-20, 22, 23].

## III. Results and Discussion

The DMI influences spin wave modes differently depending on their spin alignments. Two types of spin wave modes are investigated here: (i) Damon-Eshbach (DE) modes propagating along the $x$ axis perpendicular to the $\mathbf{H}$ direction (Fig. 1d), and (ii) backward volume (BWVM) modes propagating along the $z$ axis parallel to the $\mathbf{H}$ direction. Figures 2a, 2b show typical BLS spectra for DE spin wave modes and BWVM modes under opposite $\mathbf{H}$ directions on the Ta(5)/CoFeB(1)/Pt(0.15)/MgO thin films. In Fig. 2a, both the intensities and frequencies of the Stokes and anti-Stokes peaks are asymmetric, while such asymmetric features can be interchanged by reversing the $\mathbf{H}$ direction. The intensity asymmetry originates from the surface propagating characteristics of DE spin waves[24]: higher (lower) intensity corresponds to the ones propagating near top (bottom) surface; the propagating direction of DE spin waves ($\pm \mathbf{y} \times \mathbf{H}$) on each surface is reversed by opposite $\mathbf{H}$ directions. The frequency asymmetry is described by[19, 20, 25-27]:

$$f = \frac{\gamma}{2\pi}\sqrt{(H + Jk_M^2)(H + Jk_M^2 + 4\pi M_{\text{eff}})} - sgn(M_z)\frac{\gamma}{\pi M_S}Dk_x, \quad (1)$$

when the magnetization is along the $\mathbf{H}$ direction on the thin film plane. Here $\gamma$ is the gyromagnetic ratio, $J = \frac{2A}{M_S}$ with the exchange stiffness constant $A$ and the saturation magnetization $M_S$, $k_x$ denotes the projection of $k_M$ in the $x$ direction, $4\pi M_{\text{eff}} = 4\pi M_S - H_\perp$ represents the effective demagnetization field including the influence of interfacial magnetic anisotropy field described by $H_\perp$. The first term on the right describes the spin wave dispersion without the DMI in the ultrathin film limit, and the second term accounts for the DMI. Both $D$, $k_x$ and $4\pi M_{\text{eff}}$ can be positive or negative values in the formula. $f$ is different for $k_x < 0$ (Stokes) and $k_x > 0$ (anti-Stokes), leading to the frequency asymmetry. Specifically, the DMI reduces the energy and frequency of the spin waves with clockwise spatial chirality (e.g. $k_x > 0$ in Fig. 1d (bottom)), but increases the energy and frequency of those with counter-clockwise spatial chirality (e.g. $k_x < 0$ in Fig. 1d (top)) if looking along the vector of $\mathbf{D}$ [18]. When the magnetic field is reversed, the direction of $\mathbf{D}$ remains the same but the spatial chirality of spin waves interchange in $k$ space between $k_x < 0$ (Stokes) and $k_x > 0$ (anti-Stokes). Thus the asymmetry in the spin wave dispersion is reversed along $k_x$ direction as illustrated in Fig. 1d and demonstrated in Fig. 2a. In contrast, such intensity and frequency asymmetries are not observed for the BWVM spin waves shown in Fig. 2b, owing to their bulk characteristics and the lack of the spatial chiralities $(-\mathbf{D} \cdot (\mathbf{m}_i \times \mathbf{m}_j) \neq 0)$ present in the DE modes. We note that the observed frequency asymmetry in Fig. 2a is not due to the non-reciprocity of the DE spin waves at the two surfaces[24, 28, 29] because $|k_x t(CoFeB)| \ll 1$, as discussed in Refs. [16, 18]. Instead, it can be explained by the interfacial DMI.

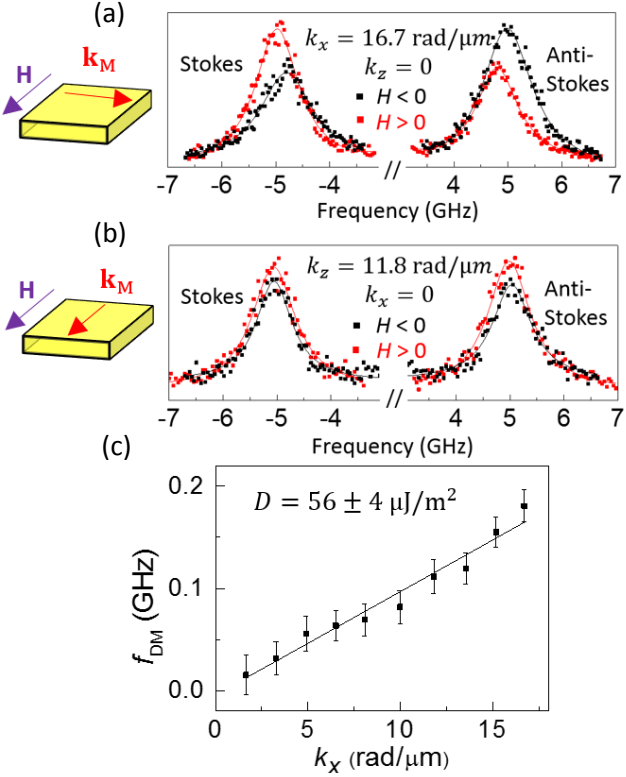

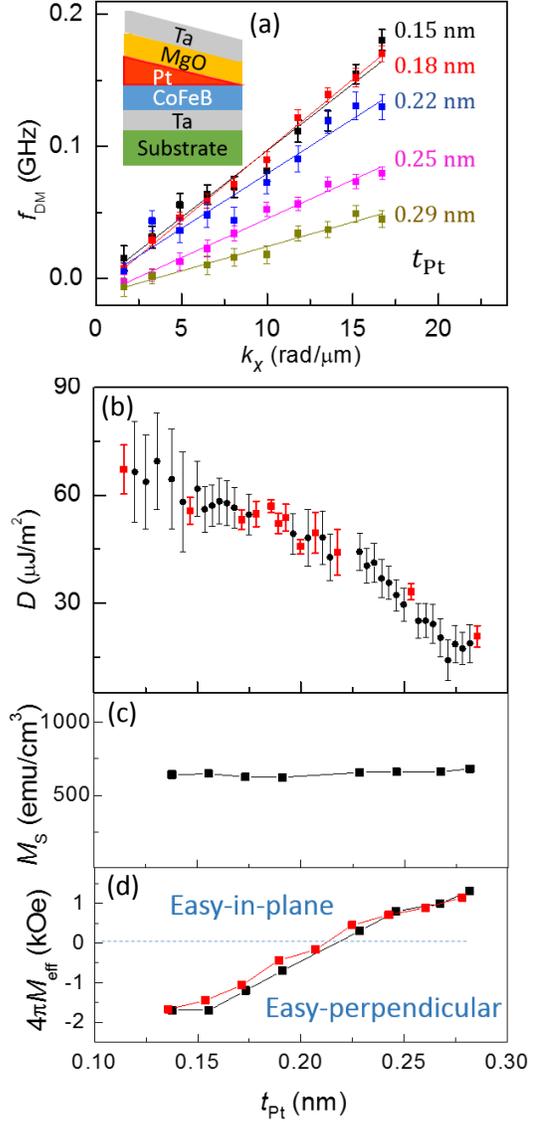

Fig. 2. BLS spectra for spin waves of (a) DE mode and (b) BWVM mode recorded at fixed incident angles under oppositely oriented external magnetic fields **H**. The solid lines represent fittings with Lorentzian functions. (c) The linear dependence of $f_{DM}$ on $k_x$. The error bars on $f_{DM}$ are determined from the deviation of Lorentzian fitting on BLS spectra, and the solid line refers to the least square fit. The uncertainty of $D$ is estimated from the deviation of the least square fit.

In order to quantify the interfacial DMI constant $D$, momentum resolved BLS measurements were performed through varying the incident angle[30]. According to Equation 1, a linear correlation exists between $f(-k_x) - f(k_x)$ and $k_x$, where the slope is determined by $D$[23, 31]. To further avoid possible instrument frequency offsets in the frequency difference between Stokes and anti-Stokes peaks, we subtract the values obtained by reversing the applied **H** field as described by the following equation.

$$f_{DM} = \frac{\left( (f(-k_x, M_z) - f(k_x, M_z)) - (f(-k_x, -M_z) - f(k_x, -M_z)) \right)}{2}$$
$$= \frac{2\gamma}{\pi M_S} D k_x \quad (2)$$

Figure 2c plots the DMI induced frequency shift $f_{DM}$ as a function of $k_x$ for the Ta(5)/CoFeB(1)/Pt(0.15)/MgO thin film, which can be linearly fitted. Considering $\gamma = 17.6$ GHz/kOe and taking $M_S = 640$ emu/cm$^3$ obtained from vibrating sample magnetometer (VSM) measurements, $D = 56 \pm 4$ μJ/m$^2$ is determined from the slope of the linear correlation. We note that the $D$ value here is modest compared with that in the Pt/FM systems[19, 20, 22, 23, 31]. Nevertheless, the DMI on the annealed Ta/CoFeB structures is still of great interest for the study of skyrmions. The enhanced interfacial magnetic anisotropy field $H_\perp$ in our sample compensates the demagnetization field in thin films and brings $4\pi M_{eff}$ close to zero, a condition favorable for establishing room temperature skyrmions[13].

Fig. 3. (a) The linear dependences of $f_{DM}$ on $k_x$ for a Pt wedge layer with a thickness gradient, where the inset illustrates the multilayer structure. The solid lines show the least square fits. (b) The interfacial DMI constant $D$ as a function of the Pt layer thickness $t_{Pt}$. The red squares are derived from the linear fits on the $k_x$ dependences of $f_{DM}$ with different $t_{Pt}$ using Equation 2. The black circles are obtainted through $f_{DM}/k_x$ at fixed $k_x = 16.7$ rad/μm by varying $t_{Pt}$ in a smaller step size. (c) $M_S$ and (d) $4\pi M_{eff}$ as functions of $t_{Pt}$, where solide lines are guides for the eye. The red (black) dots are obtained from field dependence BLS (VSM) measurements[32].

To control the DMI and to elucidate the role of the Pt layer, a wedged Pt layer was inserted between the CoFeB and MgO layers. Therefore, a spatial gradient in the DMI is introduced. The $k_x$-dependent BLS spectra at different positions on the heterostructures correspond to different Pt thicknesses (Fig. 3a). The influence of Pt-wedge on $M_s$ is negligibly small, as shown in Fig. 3c. Thus, according to Eq. 2, the observed change in the slope (Fig. 3a) mainly originates from the change in $D$ value. $D$ is found to vary by more than a factor of three when the nominal Pt thickness varies from 0.11 to 0.29 nm as shown in Fig. 3b. We note that the magnetic anisotropy also gradually changes as the Pt insertion increases as determined from field dependence BLS and VSM results[32] (Fig. 3d). This is owing to the fact that Pt atoms weaken the Co-O and Fe-O bonds at the interface between CoFeB and MgO[13]. This change in magnetic anisotropy may lead to slight underestimation in the modification of $D$ [32], and it offers another control knob for engineering skymions.

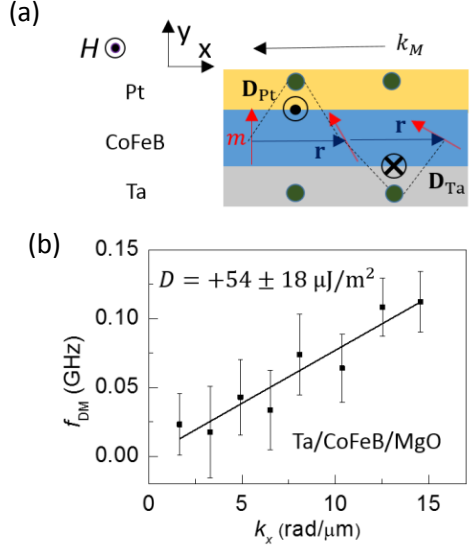

Fig. 4. (a) Schematic illustrations of directions of $\mathbf{D}_{Ta}$ (pointing along $-z$) and $\mathbf{D}_{Pt}$ (pointing along $+z$) on annealled Ta/CoFeB/Pt/MgO. The green dots represent the HM atoms. (b) The linear dependence of $f_{DM}$ on $k_x$ extracted from a control sample, annealled Ta(5)/CoFeB(1.1)/MgO.

In our sample, both the bottom Ta/CoFeB and top CoFeB/Pt interfaces contribute to DMI but partially cancel each other. The DMI vector for two spins with spatial displacement $\mathbf{r}$ can be written as $\mathbf{D}_{Ta(Pt)} = D_{Ta(Pt)} \mathbf{n}_{Ta(Pt)} \times \mathbf{r}/r$, where $\mathbf{n}_{Ta(Pt)}$ denotes the unit vector pointing from Ta(Pt) to CoFeB and $\mathbf{n}_{Ta} = -\mathbf{n}_{Pt}$, as illustrated in Figure 4(a). Although the spin Hall angles of Ta and Pt are opposite, the $D_{Ta(Pt)}$ induced by Pt and Ta on ferromagnetic metals have the same sign[2, 21]. Assuming that $\mathbf{r} = \mathbf{x}$ and $D_{Ta(Pt)} > 0$, $\mathbf{D}_{Ta(Pt)}$ induced from the bottom Ta (top Pt) layer lies in $-z$ ($+z$) direction owing to the opposite directions of $\mathbf{n}_{Ta(Pt)}$ (Fig. 4a). As the Pt layer becomes thicker, more Pt atoms contribute to DMI, and hence partially compensate the DMI initially established by the bottom Ta layer. Moreover, a very small amount of Pt is sufficient to effectively compensate the DMI introduced by the Ta layer because the $D_{Pt}$ induced by Pt is much stronger than the $D_{Ta}$ by Ta[1, 2, 4, 22].

Next, we discuss that the positive sign of $D$ found in our Ta(5)/CoFeB(1)/Pt(wedge)/MgO structures may be related to specific atomic arrangements at the HM/FM interface. In order to exclude the impact of the Pt layer in Fig. 3a, BLS studies were performed on a control sample Ta(5)/CoFeB(1.1)/MgO with the same growth condition. Figure 4b shows that this particular control sample Ta(5)/CoFeB(1.1)/MgO also exhibits a positive $D$ (right-handed magnetic chirality), whereas negative $D$ values have been reported by other groups in similar HM/FM structures, such as Pt/CoFe and Pt/Co.[2, 14, 19]. We speculate that the diffusion of B atoms in CoFeB during annealling causes the change of $D$. It has been shown that B atoms diffuse towards the interface during the annealing procedure and modify the relative positions of FM and HM atoms at CoFeB/heavy metal interfaces[21], which in turn is expected to lead to a modified DMI. In addition, a strong accumulation of B in the bottom Ta layer may affect the electronegativity of the heavy metal layer, and hence reverse the sign of $D$. Our speculation is supported by previous studies in which nitrogen-doped-Ta/CoFeB structure is shown to exhibit opposite DMI constants compared with pure-Ta/CoFeB system[4]. A similar effect has also been observed on annealed Ta/CoFeB and Pt/CoFeB structures through the domain wall studies by us and others[21, 33], where $D$ becomes positive.

Finally, we note that the $D$ value of Ta(5)/CoFeB(1.1)/MgO sample in Fig. 4b is slightly smaller than that of Ta(5)/CoFeB(1)/Pt(wedge)/MgO in Fig. 3b in the limit of zero Pt thickness. There are some variations between the samples that contributed to this apparent difference. First the thickness of the magnetic layer (CoFeB) is different, leading to an expected reduction in $D$ for the control sample Ta(5)/CoFeB(1.1)/ MgO. Previous studies have shown that $D$ is roughly linear dependent on $1/t_{FM}$ despite some deviation in the ultrathin region[17, 18]. In addition, there might be small variance in $D$ values among different samples, as it is highly sensitive to the interface quality[22].

## V. CONCLUSION

In conclusion, we demonstrate effective interfacial control of DMI on annealed Ta/CoFeB/(Pt)/MgO multilayer thin films, where the overall DMI strength results from the additive effect of both the Ta/CoFeB and CoFeB/Pt interfaces. Continuous tuning of the DMI constant $D$ by more than a factor of three is realized via less than one monolayer of Pt insertion in between the CoFeB and MgO layers. The larger net $D$ occurs at positions with the thinner Pt-wedge and the final sign of $D$ is determined by the Ta layer. Our results demonstrate that the use of two HM, DMI-active layers provides an efficient way of DMI control in the magnetic multilayers. Our work suggests that simultaneous enhancement of $D$ and the reduction of magnetic anisotropies can be realized by choosing HM materials and their thicknesses properly. This flexibility in materials properties engineering may enable skyrmions with nanometer dimensions at room temperature, which is highly desirable for high density spintronic applications.

## ACKNOWLEDGEMENTS

The work at UT-Austin and UCLA are supported by SHINES, an Energy Frontier Research Center funded by the U.S. Department of Energy (DoE), Office of Science, Basic Energy Science (BES) under award # DE-SC0012670. The work at University of Delaware is supported by NSF DMR grant #1505192.

## Reference

[1] K.-S. Ryu, L. Thomas, S.-H. Yang, and S. Parkin, Nat Nano **8**, 527 (2013).

[2] S. Emori, U. Bauer, S.-M. Ahn, E. Martinez, and G. S. D. Beach, Nat Mater **12**, 611 (2013).

[3] G. Chen, T. Ma, A. T. N'Diaye, H. Kwon, C. Won, Y. Wu, and A. K. Schmid, Nat Commun **4** (2013).

[4] J. Torrejon, J. Kim, J. Sinha, S. Mitani, M. Hayashi, M. Yamanouchi, and H. Ohno, Nat Commun **5** (2014).

[5] S. Pizzini, et al., Physical Review Letters **113**, 047203 (2014).

[6] N. Romming, C. Hanneken, M. Menzel, J. E. Bickel, B. Wolter, K. von Bergmann, A. Kubetzka, and R. Wiesendanger, Science **341**, 636 (2013).

[7] N. Nagaosa and Y. Tokura, Nat Nano **8**, 899 (2013).

[8] T. Schulz, et al., Nat Phys **8**, 301 (2012).

[9] W. Jiang, et al., Science **349**, 283 (2015).

[10] J. Iwasaki, M. Mochizuki, and N. Nagaosa, Nat Nano **8**, 742 (2013).

[11] A. Fert, V. Cros, and J. Sampaio, Nat Nano **8**, 152 (2013).

[12] A. Fert and P. M. Levy, Physical Review Letters **44**, 1538 (1980).

[13] G. Yu, et al., Nano Letters **16**, 1981 (2016).

[14] S. Woo, et al., Nat Mater **15**, 501 (2016).

[15] G. Chen, A. Mascaraque, apos, A. T. Diaye, and A. K. Schmid, Applied Physics Letters **106**, 242404 (2015).

[16] O. Boulle, et al., Nat Nano **11**, 449 (2016).

[17] J. Cho, et al., Nat Commun **6** (2015).

[18] H. T. Nembach, J. M. Shaw, M. Weiler, E. Jue, and T. J. Silva, Nat Phys **11**, 825 (2015).

[19] M. Belmeguenai, J.-P. Adam, Y. Roussigné, S. Eimer, T. Devolder, J.-V. Kim, S. M. Cherif, A. Stashkevich, and A. Thiaville, Physical Review B **91**, 180405 (2015).

[20] K. Di, V. L. Zhang, H. S. Lim, S. C. Ng, M. H. Kuok, J. Yu, J. Yoon, X. Qiu, and H. Yang, Physical Review Letters **114**, 047201 (2015).

[21] R. Lo Conte, et al., Physical Review B **91**, 014433 (2015).

[22] N.-H. Kim, D.-S. Han, J. Jung, J. Cho, J.-S. Kim, H. J. M. Swagten, and C.-Y. You, Applied Physics Letters **107**, 142408 (2015).

[23] K. Di, V. L. Zhang, H. S. Lim, S. C. Ng, M. H. Kuok, X. Qiu, and H. Yang, Applied Physics Letters **106**, 052403 (2015).

[24] R. E. Camley, Surface Science Reports **7**, 103 (1987).

[25] J.-H. Moon, S.-M. Seo, K.-J. Lee, K.-W. Kim, J. Ryu, H.-W. Lee, R. D. McMichael, and M. D. Stiles, Physical Review B **88**, 184404 (2013).

[26] S. Rohart and A. Thiaville, Physical Review B **88**, 184422 (2013).

[27] A. N. Bogdanov and U. K. Rößler, Physical Review Letters **87**, 037203 (2001).

[28] G. Chen, et al., Physical Review Letters **110**, 177204 (2013).

[29] M. Suzuki, et al., Physical Review B **72**, 054430 (2005).

[30] Z. K. Wang, V. L. Zhang, H. S. Lim, S. C. Ng, M. H. Kuok, S. Jain, and A. O. Adeyeye, Applied Physics Letters **94**, 083112 (2009).

[31] V. L. Zhang, K. Di, H. S. Lim, S. C. Ng, M. H. Kuok, J. Yu, J. Yoon, X. Qiu, and H. Yang, Applied Physics Letters **107**, 022402 (2015).

[32] See supplementary information for detailed results and analysis on field dependence BLS and VSM.

[33] G. Yu, P. Upadhyaya, K. L. Wong, W. Jiang, J. G. Alzate, J. Tang, P. K. Amiri, and K. L. Wang, Physical Review B **89**, 104421 (2014).